\documentclass{article}

\usepackage[utf8]{inputenc}
\usepackage{graphicx}
\usepackage{amsmath}
\usepackage{amsfonts}
\usepackage{amssymb}
\usepackage{authblk}
\usepackage{caption}
\usepackage{hyperref}
\usepackage{ragged2e}
\usepackage{setspace}
\usepackage[left=1in, right=1in, top=1in, bottom=1in]{geometry}

\title{Switchable Exchange Bias Resulting from Correlated Domain Structures in Orthogonally Coupled Antiferromagnet/Ferromagnet van der Waals Heterostructures}

\author[1]{Aditya Kumar}
\author[1]{Sadeed Hameed}
\author[2]{Thibaud Denneulin}
\author[1,]{Aravind Puthirath Balan}
\author[2]{Joseph Vas}
\author[1]{Kilian Leutner}
\author[3]{Lei Gao}
\author[1]{Olena Gomonay}
\author[1]{Jairo Sinova}
\author[2]{Rafal E. Dunin-Borkowski}
\author[1,4,]{Mathias Kläui}

\affil[1]{Institute of Physics, Johannes Gutenberg University Mainz, Mainz, Germany}
\affil[2]{Ernst Ruska-Centre for Microscopy and Spectroscopy with Electrons, Forschungszentrum Jülich, Jülich, Germany}
\affil[3]{Max Planck Institute for Polymer Research Mainz, Mainz, Germany}
\affil[4]{Centre for Quantum Spintronics, Department of Physics, Norwegian University of Science and Technology, Trondheim, Norway}
\affil[*]{Email: aravindputhirath@uni-mainz.de, klaeui@uni-mainz.de}

\date{}

\begin{document}

\maketitle
\begin{justify}

\doublespacing

\begin{abstract}
Van der Waals (vdW) magnetic heterostructures offer a versatile platform for engineering interfacial spin interactions with atomic precision, enabling nontrivial spin textures and dynamic behaviors. In this work, we report robust asymmetric magnetization reversal and exchange bias in Fe\(_3\)GeTe\(_2\) (FGT), driven by interlayer exchange coupling with the A-type antiferromagnet CrSBr. Despite the orthogonal magnetic anisotropies—out-of-plane easy axis in FGT and in-plane in CrSBr—we observe a strong interfacial exchange interaction that gives rise to pronounced and switchable exchange bias and asymmetric switching in FGT, persisting up to the N\'eel temperature of CrSBr (\(\sim132\)\,K) as revealed by anomalous Hall effect measurements.

We uncover the microscopic origin of this behavior through cross-sectional magnetic imaging of the domain structure using off-axis electron holography. The results reveal that the asymmetric switching and exchange bias arise from the influence of CrSBr on the domain configuration of FGT, where the in-plane antiferromagnetic state of CrSBr promotes the formation of stripe-like domain structures in FGT with circular rotation of magnetization in the cross-sectional bc plane defined by the easy axes of both FGT and CrSBr. These findings elucidate the mechanism of exchange bias in orthogonally coupled van der Waals systems and demonstrate a pathway for stabilizing three-dimensional domain structures in ferromagnets through interfacial exchange interactions.

\end{abstract}


\section{Introduction}

The emergence of van der Waals (vdW) magnetic materials has opened new avenues for investigating spin-dependent interactions and advancing the development of next-generation spintronic devices. Due to their intrinsically layered atomic structure, vdW materials facilitate the assembly of magnetic heterostructures with atomic-level precision, free from the constraints of lattice mismatch and interlayer chemical bonding.\cite{Novoselov2016, Burch2018} Such heterostructures offer a highly adaptable platform for engineering magnetic properties, including magnetic anisotropy and interlayer exchange coupling. As a result, vdW magnetic systems provide fertile ground for exploring interfacial magnetic phenomena such as spin reorientation transitions, tunneling magnetoresistance, and exchange bias effects.\cite{bera2023unravelling,Song2019, Zhao2025,Zhang2022_SRT, Zhang2024,Chen2024, Cheng2024, phan2023exchange, Balan2024_ACSNano, Balan2024_AdvMater}

The advantages offered by vdW materials in device fabrication and manipulation have fueled significant advances in the field of two-dimensional (2D) materials, with magnetic vdW systems becoming a very active area of research. The expanding library of magnetic vdW compounds now includes a diverse range of out-of-plane (OOP) and in-plane (IP) magnetized ferromagnets and antiferromagnets.\cite{gibertini2019magnetic} Ferromagnets such as Fe\(_3\)GaTe\(_2\), which exhibit Curie temperatures ($T_{C}$) well above room temperature, have recently been identified.\cite{wu2024robust} In parallel, advances in wafer-scale semi-automated transfer techniques and scalable deposition methods have further reinforced the potential of vdW materials for integration into spintronic applications.\cite{liu2023wafer, li2024wafer} These developments point toward the increasing feasibility of incorporating vdW magnets into future spintronic device architectures.

Exchange bias has long served as a foundational mechanism in magnetic memory technologies, particularly in stabilizing the reference (fixed) layer of magnetic tunnel junctions.\cite{Meiklejohn1956,kools2002exchange} This interfacial phenomenon, primarily observed in antiferromagnet/ferromagnet (AFM/FM) bilayer systems, manifests as a horizontal shift in the magnetic hysteresis loop and arises from exchange coupling between uncompensated antiferromagnetic spins and adjacent ferromagnetic moments. Traditionally, exchange bias has been extensively studied in thin film metallic multilayers such as Co/IrMn and NiFe/FeMn, where the effect is influenced by factors such as interface roughness and grain boundaries.\cite{Nogues1999, stamps2000mechanisms} In contrast, vdW magnetic heterostructures provide atomically sharp interfaces, minimal interdiffusion, and improved control over both dimensionality and interfacial quality. These features establish a cleaner and more tunable platform for realizing and manipulating exchange bias effects.\cite{phan2023exchange}

In conventional exchange bias systems, collinear alignment between the anisotropies of the FM and AFM—typically both IP or both OOP—has been extensively utilized to achieve large unidirectional anisotropies, which are critical for applications in spin valves and magnetic tunnel junctions. \cite{kools2002exchange} However, recent efforts have focused on systems with orthogonal magnetic anisotropies, in which the two layers possess mutually perpendicular easy axes, such as OOP FM coupled to IP FM or AFM order. Such orthogonal coupling at the interface breaks interfacial rotational symmetry and can stabilize canted spin states, chiral domain walls, and topologically non-trivial spin textures.\cite{fernandez2019symmetry, avci2021chiral} These effects not only facilitate deterministic field-free switching via symmetry breaking,\cite{oh2016field, Cham2024} but also open new avenues for domain nucleation and stabilization.\cite{yu2018room} Elucidating the interfacial spin interactions in such non-collinear systems is therefore vital for advancing the design of topologically robust, energy-efficient magnetic devices.

Exchange bias in vdW heterostructures has been demonstrated in a range of collinear AFM/FM heterostructures comprising the OOP ferromagnet Fe\(_3\)GeTe\(_2\) (FGT) coupled with many OOP AFMs including MnPSe\(_3\), CrPS\(_4\) and MnPS\(_3\).\cite{FGT_nat_comm, Dai2021, Balan2024_ACSNano, Balan2024_AdvMater} In contrast, the choice of IP antiferromagnets remains limited. Reported examples include CrCl\(_3\),\cite{cai2019atomically} NiPS\(_3\),\cite{kang2020coherent} CoPS\(_3\),\cite{liu2021magnetic} and CrSBr.\cite{lopez2022dynamic} Among these, NiPS\(_3\) and CoPS\(_3\) exhibit fully compensated spin structures within individual layers, which is generally unfavorable for achieving significant interfacial exchange coupling.\cite{stamps2000mechanisms, schippers2020large}

CrCl\(_3\), while of interest,\cite{Zhu2020, bedoya2021intrinsic} in addition to a low transition temperature, also suffers from poor environmental stability, which increases the risk of interfacial intermixing similar to effects observed in MnPS\(_3\)/FGT systems.\cite{Balan2024_AdvMater, Mastrippolito2021} In contrast, CrSBr demonstrates robust chemical stability under ambient conditions,\cite{Tschudin2024} and has recently been shown to form pristine, atomically sharp interfaces with FGT.\cite{Ziebel2024, Cham2024} Moreover, the weak interlayer antiferromagnetic coupling in CrSBr enables a tunable transition to a ferromagnetic configuration under modest external magnetic fields.\cite{long2023intrinsic} This can enable preset-field-induced, easily switchable exchange bias, as reported recently in FM/Ferrimagnet systems.\cite{Balan2024_ACSNano, radu2012perpendicular} The combination of chemical stability, interfacial abruptness, and magnetic tunability establish CrSBr/FGT as a promising model system for probing the microscopic mechanisms underlying exchange bias and domain evolution in orthogonally coupled magnetic heterostructures. 

Recently, magnetotransport measurements on such orthogonally coupled CrSBr/Fe\(_3\)GeTe\(_2\)/Pt trilayers have revealed IP exchange bias and an IP canting of the magnetization. Notably, no OOP exchange bias was detected in these systems.\cite{Cham2024} In a separate study, Ni \textit{et al.} investigated CrSBr/Fe\(_3\)GaTe\(_2\) heterostructures and observed interfacial exchange coupling, with measurable effects confined to temperatures below 50~K.\cite{Ni2025} The contrasting observations in these CrSBr-based systems raise important questions regarding the underlying mechanism of exchange bias and the nature of interfacial magnetic interactions in heterostructures with orthogonal spin configurations.

In this work, we investigate CrSBr/FGT vdW heterostructures that combine an IP antiferromagnet (CrSBr) with an OOP ferromagnet FGT, resulting in competing orthogonal anisotropies at the interface. We probe exchange bias in this system through temperature-dependent anomalous Hall effect (AHE) measurements, which provide insight into the magnetic hysteresis behavior of the CrSBr/FGT heterostructures. The observed exchange bias is characterized by a typical moderate shift along the field axis, along with a sizable asymmetry in an otherwise square hysteresis loop observed for the exfoliated FGT flake.\cite{FGT_nat_comm} To understand the origin of the observed exchange bias and the asymmetry in the hysteresis loops, we employ cross-sectional magnetic imaging via off-axis electron holography.\cite{Midgley2009} Based on our experimental observations, we propose a phenomenological model of exchange bias in orthogonally coupled AFM/FM systems, highlighting the role of interfacial spin configurations and the influence of weakly coupled antiferromagnetic layers.

\section{Results and Discussion}

\subsection{Structural characterization of CrSBr/FGT vdW heterostructures}

Bulk FGT single crystals are synthesized via chemical vapor transport,\cite{deiseroth2006fe3gete2} while CrSBr and hexagonal boron nitride (h-BN) crystals are commercially obtained from HQ Graphene.\cite{hqgrapheneCrSBr} CrSBr/Fe\(_3\)GeTe\(_2\) heterostructures capped with h-BN layers are fabricated by mechanical exfoliation of bulk crystals followed by a polymer-assisted dry-transfer process, as described in detail in the Experimental section. The full heterostructure—comprising FGT, CrSBr, and a protective h-BN capping layer—is transferred onto pre-patterned gold contacts in an inert environment to preserve interface quality and to prevent any surface oxidation. The CrSBr layer in contact with hBN is sufficiently thick to render any possible hybridization effects from hBN insignificant.

A schematic for a fabricated device (D1) and a corresponding optical micrograph are shown in Figure~\ref{fig:device_EB_SQUID}a. Atomic force microscopy (AFM) is used to determine the thicknesses of the exfoliated flakes. The FGT layer is measured to be approximately 87$\pm$2\,nm thick, while the CrSBr layer is around 30$\pm$2\,nm. The corresponding AFM image and line scans used for thickness estimation are provided in Figure~SI1a of the \textit{Supporting Information}. Additionally, Raman spectroscopy is performed on both FGT and CrSBr flakes to assess sample quality; the resulting spectra typical of these materials are presented in Figure~SI2.

The quality of the CrSBr/FGT interface was further assessed using cross-sectional scanning transmission electron microscopy (STEM) equipped with a high-angle annular dark-field (HAADF) detector and an energy dispersive X-ray (EDX) detector for elemental mapping of the interface. As shown in Figures~\ref{fig:device_EB_SQUID}b and c, the STEM and element-sensitive EDX mappings confirm the formation of a clean interface without any evidence of interdiffusion between the layers, as seen in MnPS\(_3\)/FGT vdW heterostructures.\cite{Balan2024_AdvMater} These results establish the structural integrity and compositional sharpness necessary to facilitate well-defined interfacial magnetic coupling in the as-fabricated heterostructures.

\subsection{Asymmetric magnetization switching and exchange bias in CrSBr/FGT vdW heterostructures}

To probe the magnetic hysteresis of the fabricated CrSBr/FGT heterostructure in D1, we employed anomalous Hall effect (AHE) measurements.\cite{nagaosa2010anomalous} Due to the small geometrical size of the heterostructure, the total magnetic moment is too low to be detected via volumetric techniques such as SQUID magnetometry. Instead of using extensive properties, techniques that probe intensive properties like the AHE have become widely adopted to characterize the magnetic hysteresis of metallic ferromagnetic flakes.

Exchange bias in thin-film AFM/FM systems is typically induced by field-cooling: the system is heated above the $T_N$ of the antiferromagnet and subsequently cooled in the presence of a magnetic field to align the spin of the AFM in a preferred orientation. This thermal activation is critical because many thin-film AFMs exhibit spin-flip fields on the order of tens of Tesla, which are beyond the reach of standard experimental setups.\cite{behovits2023terahertz} However, CrSBr exhibits weak interlayer antiferromagnetic coupling,\cite{Liu2024} resulting in a significantly reduced spin-flip field of approximately 300\,mT. Figure~\ref{fig:device_EB_SQUID}e shows bulk magnetization versus applied magnetic field measurements for CrSBr, revealing saturation around 0.3\,T along the \(b\)-axis (the magnetic easy axis), 1\,T along the \(a\)-axis, and 2\,T along the OOP (\(c\)-axis) direction. Given this low saturation field, we bypass the conventional field-cooling process and instead apply a preset-field of $\pm2.5$\,T along the OOP direction prior to hysteresis measurements to induce exchange bias. Throughout the following discussion, the direction of preset-field applied is along the OOP direction unless otherwise mentioned. 

Figure~\ref{fig:device_EB_SQUID}d displays the AHE hysteresis loops measured at 10\,K after applying preset-fields of $+2.5$\,T and $-2.5$\,T. For a $+2.5$\,T preset-field, the magnetization reversal from the positive saturated state exhibits sharp, monodomain switching, whereas the reversal on the negative side is more gradual, indicative of domain nucleation and propagation. This asymmetry is reversed when the preset-field is $-2.5$\,T. The exchange bias field $H_\mathrm{EB}$ is quantified as the average of the positive ($H_C^+$) and negative ($H_C^-$) coercive fields, defined at the points where $V_{XY}$ reaches the midpoint between the saturation voltages:
\[
H_\mathrm{EB} = \frac{H_C^+ + H_C^-}{2}.
\]
At 10\,K, we observe an exchange bias of $-47 \pm 5$\,mT and $+45 \pm 5$\,mT following $+2.5$\,T and $-2.5$\,T preset-fields, respectively. The $\pm5$\,mT uncertainty corresponds to the magnetic field step size used in the hysteresis measurements. To verify reproducibility, we perform exchange bias measurements on a second device (D2) with comparable FGT thickness, and the results are in good agreement despite the difference in CrSBr thickness, and are included in the \textit{Supporting Information} (Figure~SI1b). The hysteresis loop shape and measured exchange bias closely resemble those observed in D1, as shown in Figure~SI3 in the \textit{Supporting Information}.

We next performed AHE measurements using preset-fields of $\pm2.5$\,T across a temperature range from 10\,K to 150\,K. The resulting hysteresis loops for the $+2.5$\,T case are shown in Figures~\ref{fig:fig2}a and b. Corresponding plots for the $-2.5$\,T preset-field are similar but with opposite bias direction and are provided in \textit{Supporting Information} Figures~SI4a,b. The extracted exchange bias values as a function of temperature are plotted in Figure~\ref{fig:fig2}c. The temperature dependence of coercivity of this vdW heterostructure can be found in \textit{Supporting Information} Figure~SI4c. We find that the exchange bias decreases monotonically, vanishing near 130\,K—close to the reported $T_N \approx 132$\,K of CrSBr.\cite{long2023intrinsic} 

These observations differ substantially from the results reported for the CrSBr/FGT/Pt trilayer by Cham \textit{et al.}\cite{Cham2024} and the CrSBr/Fe$_3$GaTe$_2$ vdW heterostructure by Ni \textit{et al.}\cite{Ni2025}. 

Cham \textit{et al.} did not detect any OOP exchange bias in the CrSBr/FGT/Pt trilayer. Two key distinctions between the two systems may account for this discrepancy. First, Cham \textit{et al.} employed much thinner FGT flakes (9--12\,nm), while our study utilizes thicker flakes, which exhibit reduced effective OOP anisotropy and a stronger tendency towards domain nucleation.\cite{birch2022history} Second, their device includes a Pt underlayer, which may alter the magnetic anisotropy of FGT through interfacial interactions similar to those observed in FGT/WSe\(_2\) heterostructures.\cite{kim2021interface} These differences in thickness, anisotropy, and device composition contribute to structural and magnetic disparities between the systems, and can account for the observed differences in exchange bias and hysteresis loop characteristics.

Ni \textit{et al.} reported the existence of exchange bias in CrSBr/Fe$_3$GaTe$_2$ vdW heterostructures. However, the observed blocking temperature ($T_B$) is significantly lower than that exhibited by the CrSBr/FGT system studied in this work. The different anisotropy and exchange coupling of FGT and Fe$_3$GaTe$_2$ could account for this behavior. Moreover, they do not discuss a pronounced training effect characterized by the disappearance of exchange bias in subsequent hysteresis loops. 

Zero OOP exchange bias observed in thinner FGT flakes,\cite{Cham2024} and significant exchange bias in thicker flakes as observed in our study further highlights the complex, non-monotonic thickness dependence of exchange bias in vdW heterostructures. Typically, the exchange bias is expected to decrease inversely with the thickness of the ferromagnetic layer.\cite{binek2001exchange} In vdW heterostructures with weakly coupled antiferromagnets, the thickness dependence of exchange bias is often complex and does not always follow a simple monotonic scaling.\cite{albarakati2022electric}

The exchange bias in this CrSBr/FGT vdW heterostructure emerges through monodomain switching on one side, whereas a gradual, domain-mediated reversal on the other, rather than manifesting as a rigid horizontal shift. As shown in Figure~\ref{fig:fig2}d, for a $+2.5$\,T preset-field (corresponding data for a $-2.5$\,T preset-field can be found in \textit{Supporting Information} Figure~SI4d), $H_C^-$ (corresponding to monodomain switching) mirrors the temperature dependence of the exchange bias, whereas $H_C^+$ (gradual switching) follows the typical coercivity vs. temperature trend of a ferromagnet.\cite{FGT_nat_comm} The dependence of how exchange bias varies as a function of preset-field is also provided in \textit{Supporting Information} Figure~SI4e. The similar trend of $H_C^-$ and $H_{EB}$ as a function of temperature are plotted in Figure~SI5 to highlight the influence of the field corresponding to monodomain switching on exchange bias.This suggests that the preset-field induces an abrupt switching of FGT moments in one direction, while magnetization reversal in the opposite direction is gradual, driven by domain formation and domain wall propagation, thereby generating a sizable exchange bias. Training effect measurements, shown in Figure~\ref{fig:fig2}e, indicate that the first hysteresis loop following the preset-field exhibits pronounced asymmetry and clear exchange bias. All subsequent loops, however, are symmetric and display gradual switching on both sides, which is unusual considering the abruptness of the disappearance of exchange bias.

While the intuitive models proposed by Liu \textit{et al.}\cite{Liu2024} and Ni \textit{et al.}\cite{Ni2025} qualitatively explain the observation of asymmetric hysteresis, they do not fully capture the details of the experimental behavior observed in this work. Liu \textit{et al.}\cite{Liu2024} investigate Fe\textsubscript{3}GaTe\textsubscript{2}/O-Fe\textsubscript{3}GaTe\textsubscript{2}, which exhibit a collinear OOP spin configuration. Ni \textit{et al.} address asymmetric switching in a non-collinear antiferromagnet; however, their model cannot explain the pronounced training effect observed in our experiments.

To understand the underlying mechanism for this behavior, we perform cross-sectional off-axis electron holography on the CrSBr/FGT vdW heterostructure, and develop a phenomenological model describing the exchange coupling and domain wall evolution, as discussed next.

\subsection{Understanding asymmetric switching mechanism and exchange bias in CrSBr/FGT vdW heterostructure}

The characteristic asymmetric hysteresis loop following a preset-field and its disappearance in subsequent cycles as discussed in the previous section suggests the significant role of a preset-field induced metastable state,\cite{egorov2025metastable} in CrSBr. Such a fully OOP polarized metastable state facilitates monodomain switching in the opposite direction to the applied preset-field. The gradual switching via domain nucleation and domain wall propagation observed in magnetization reversal and the symmetric switching observed in the subsequent loops indicates that this metastable state has been disappeared during the initial magnetic reversal after the application of a preset-field, which results in zero exchange bias in the following field cycles. The minor loop measurements performed with and without the application of a preset-field (shown in \textit{Supporting Information} Figure~SI6) further corroborate these findings. 

During domain nucleation and domain wall propagation, the spins in CrSBr are expected to align along their magnetic easy axis—the \(b\)-axis—while the OOP anisotropy of FGT favors spin alignment along the \(c\)-axis.

 In this configuration, the magnetization profile of FGT, particularly near the interface, is governed by the competition among three key interactions across its thickness: First, interfacial exchange coupling promotes parallel spin alignment between FGT and CrSBr. Second, the intrinsic anisotropies of the two materials—OOP for FGT and IP for CrSBr—favor orthogonal spin orientations. Third, dipolar interactions lead to the formation of flux closure domains, thereby favoring aligned interfacial spins.\cite{labrune2001micromagnetic,hermosa20223d}
The OOP anisotropy (\(K_c\)) of CrSBr, estimated from antiferromagnetic resonance measurements in a recent study, is approximately 0.255\,MJ/m\textsuperscript{3} at 5\,K.\cite{cham2022anisotropic} In contrast, the uniaxial anisotropy energy of an FGT flake has been reported as 1.46\,MJ/m\textsuperscript{3} at a similar temperature.\cite{leon2016magnetic}
 Despite the significantly larger anisotropy of FGT, the difference is compensated by the dipolar interactions in the thick FGT flake, which favors IP alignment of interfacial spins.
  Given that the CrSBr has an IP easy axis along $b$, the rotation of spins in a closure domain would take place along the $bc$ plane. 
  
  To study this interplay in detail, it is necessary to resolve the magnetization profile of the AFM/FM heterostructure along the $bc$ cross-section with nanometer-scale spatial resolution. This can be achieved with off-axis electron holography, a powerful transmission electron microscopy (TEM) technique capable of resolving magnetic structures with spatial resolution on the order of 10\,nm.\cite{Midgley2009,denneulin2025acquisition} For this measurement, a cross-section lamella of the CrSBr/FGT vdW heterostructure is fabricated using focused ion beam (FIB) milling along the $bc$-plane of CrSBr. The lamella is cooled to 95\,K to induce magnetic ordering and is measured at zero field after application of 1\,T of magnetic field. The 1\,T field used in holography is lower than the 2.5\,T preset field applied during transport measurements. However, the preset-field dependence of exchange bias (Figure~SI4e) shows no significant change in exchange bias above 1\,T, making the results directly comparable.
Figure~\ref{fig:fig3}a shows the raw phase map obtained at 95\,K and 0\,mT with the corresponding phase gradient image shown in \textit{Supporting Information} Figure~SI7d. The observed phase map shows no inhomogeneous magnetization, indicating a saturated monodomain state in FGT and CrSBr. However, measuring a demagnetized state as shown in Figure~\ref{fig:fig3}b reveals the presence of magnetic domains in FGT.

To further resolve the internal domain structure, we processed the phase map obtained at 95\,K in a demagnetized state. A background phase image acquired at 300\,K (above the $T_{C}$) is subtracted from the 95\,K phase map (Figure~\ref{fig:fig3}b) to eliminate nonmagnetic contributions. The detailed data processing steps are explained in \textit{Supporting Information} Figure~SI8. The resultant magnetic phase image is shown in Figure~\ref{fig:fig3}c. The phase gradient is calculated and color-coded to visualize the magnetic induction field direction, as shown in Figure~\ref{fig:fig3}d. 

By comparing the observed phase map with calculations performed by Hu et al., we can identify the magnetization rotation near the top and bottom surfaces of the FGT layer, forming Néel-type flux closure caps.\cite{hu2018direct} Meanwhile, the central region of the film hosts Bloch-like domain walls, with their cores magnetized along the \(a\)-axis. This together forms circular magnetic closure domains along the $bc$-plane as observed in the field map in Figure~\ref{fig:fig3}d. 

The uniform focal depth of the observed circular closure domains in our system suggests the presence of lateral stripe domains extending along the \(a\)-axis. Similar magnetic textures were recently reported in Gd\(_{12}\)Co\(_{88}\)/Nd\(_{17}\)Co\(_{83}\)/Gd\(_{24}\)Co\(_{76}\) ferrimagnetic trilayers, where orthogonally coupled interfaces lead to the formation of related spin structures.\cite{hermosa20223d}

The field evolution of such domains to a saturated state proceeds via domain wall propagation, manifesting as gradual switching on the positive side of the AHE curves for a positive preset-field, in line with our AHE results as shown in Figures~\ref{fig:fig2}a and b. The occurrence of monodomain switching in one direction and gradual reversal through domain nucleation and domain wall propagation gives rise to the distinctive asymmetric magnetization switching as observed in our CrSBr/FGT vdW heterostructures. This results in sizable exchange bias. The disappearance of both exchange bias and switching asymmetry near the \(T_\mathrm{N}\) of CrSBr (Figures~\ref{fig:fig2}c and d), along with the pronounced training effect (Figure~\ref{fig:fig2}e), and the direct observation of characteristic cross-sectional domains through electron holography, collectively support a scenario in which interfacial exchange coupling—together with the competing anisotropies of CrSBr and FGT—governs the magnetization reversal dynamics and the emergence of exchange bias.

To understand the stabilization of the metastable fully saturated ferromagnetic state of CrSBr and the AFM's role in domain nucleation, we simulate a model based on the minimization of the free energy of the AFM/FM bilayer. The details can be found in \textit{Supporting Information} section~SI5 and Figure~SI11. The stabilization of a metastable state can be attributed, possibly, to a fourth-order anisotropy term in CrSBr. This higher-order anisotropy term, along with an exchange coupling field from the FM, may enable the stabilization of a metastable ferromagnetic state in CrSBr following the application of a large preset-field. An OOP hysteresis of CrSBr at 50 K with this 4th-order anisotropy term is illustrated in Figure~SI10 as indicated by the red curve in comparison to the corresponding data for a bulk crystal (blue color). 
The preset field dependence presented in Figure~SI4e of the Supporting Information reveals that the exchange bias saturates with an applied field of 1\,T, which is considerably lower than the OOP saturation field of CrSBr. This observation provides further evidence that the exchange field from FGT reduces the effective OOP saturation field required for CrSBr. The model also supports the proposed mechanism in which interfacial exchange coupling between FGT and CrSBr promotes domain formation in FGT and induces a canted magnetization in CrSBr. This canted magnetization should not be confused with spin-flop canting, which involves a rotation of the N\'eel vector. In contrast, the observed OOP canting in CrSBr preserves the N\'eel vector along the \textit{b}-axis. In a free-standing FM layer, OOP magnetization generates strong stray fields, resulting in a large stray field energy, which can be reduced through the formation of 180° domains. However, this multidomain configuration incurs an energy cost associated with domain wall formation. In a freestanding FM layer, the domain wall energy can be high enough to suppress spontaneous domain formation. In contrast, in a FM/AFM heterostructure, interfacial exchange coupling can significantly lower the domain wall energy barrier, enabling domain wall formation and facilitating the stabilization of multidomain states. The interfacial exchange coupling energy also contributes to an increased saturation field in the symmetric loops, as observed in loops 2--4 in Figure~\ref{fig:fig2}e. This is because a larger magnetic field is required to align the interfacial spins of FGT along the OOP direction. The results show how interfacial interactions facilitate flux closure and promote domain configurations that are otherwise unfavorable in isolated FM systems. A schematic and a calculated magnetization distribution in both layers are shown in Figure~SI11a and b in the \textit{Supporting Information SI5}.

To investigate the evolution of flux-closure domains toward the saturated state, we performed history-dependent imaging measurements on the CrSBr/FGT vdW heterostructure. The measurement protocol starts with zero-field-cooling (ZFC) to 95\,K, followed by imaging in this initial state (Figure~\ref{fig:fig4}a). The resulting image reveals stripe-like flux-closure domains, similar to those observed in Figure~\ref{fig:fig3}d. Subsequently, the sample was tilted by 45° and subjected to a magnetic field with a 50\,mT OOP component. After removing the field, the sample was returned to its original orientation for imaging (Figure~\ref{fig:fig4}b). This procedure was repeated for additional OOP fields of $-50$\,mT, 75\,mT, and $-75$\,mT, producing the domain structures shown in Figures~\ref{fig:fig4}c--e. These sequential images capture the domain evolution as the system approaches the saturated state.

This history-dependent imaging clearly demonstrates the FGT magnetization process via domain evolution. A comparison between Figures~\ref{fig:fig4}a and \ref{fig:fig4}b shows the asymmetric growth of circular domains, characterized by an increase in upward-pointing spins and a reduction in downward-pointing spins. A similar trend is evident in Figures~\ref{fig:fig4}c and \ref{fig:fig4}d, until FGT reaches a monodomain state in Figure~\ref{fig:fig4}e.

It is also important to note that the domains in Figure~\ref{fig:fig4}c display different local magnetization rotation, indicating that they are regular stripe domains rather than chiral. The interfacial spins in FGT and CrSBr are parallel; however, due to the presence of multiple domains in CrSBr, local variations in interfacial exchange may induce distinct magnetization rotation in different regions of FGT.

Based on the results discussed in this section, the behavior of such an orthogonally coupled system can be interpreted through the following phenomenological model, which is summarized in Figure~\ref{fig:fig5}.

\begin{enumerate}
    \item[(i)] Application of a 2.5\,T OOP preset-magnetic field fully saturates both CrSBr and FGT along the $c$-axis. The application of a large magnetic field, combined with exchange coupling from FGT, drives CrSBr into a ferromagnetic metastable state.

    \item[(ii)] Upon removal of the preset-field, the interfacial layers of CrSBr and FGT remain aligned along the OOP direction, indicating that CrSBr retains a metastable ferromagnetic state. The absence of contrast in off-axis electron holography at zero field—following the application of a 1\,T magnetic field—confirms the OOP-saturated state of both FGT and CrSBr. 

    \item[(iii)] As a sufficiently strong negative magnetic field is applied, the bulk of the FGT layer undergoes abrupt magnetization reversal. However, CrSBr is no longer OOP-saturated and instead adopts a canted magnetic state, with the N\'eel vector aligned along the \(b\)-axis and a canted magnetization component oriented along the \(c\)-axis. The layers of CrSBr closest to the interface retain an OOP magnetization, and layers further away from the interface gradually rotate toward the IP direction.

    \item[(iv)] The magnetization reversal process—from negative to positive field—is gradual and dominated by domain nucleation (as directly observed and shown in Figure~\ref{fig:fig3}d), domain wall propagation, and annihilation. At this state, the canted magnetization in CrSBr follows the domain structure of the underlying FGT layer.

    \item[(v)] When FGT saturates again in the positive direction in the presence of a sufficiently large magnetic field, CrSBr forms a domain structure along the $c$-axis rather than saturating. The subsequent loops measured after this state cycle between states (iii), (iv), and (v), without entering states (i) and (ii), resulting in gradual switching on both sides of the hysteresis loop with zero exchange bias. 

\end{enumerate}

This model explains the asymmetry, training effect, and domain dynamics observed in both transport and imaging experiments, providing a coherent explanation for the non-trivial magnetization behavior in CrSBr/FGT heterostructures. Although domain evolution can lead to asymmetric magnetization reversal, alternative mechanism may also contribute. In thinner ferromagnetic flakes, where domain nucleation is less favorable, asymmetric reversal can result from incoherent rotation of bulk and interfacial spins, as suggested by observations in Ni \textit{et al}.\cite{Ni2025}

The 3D domains observed in the cross-section (bc plane) of CrSBr/FGT heterostructures invite further investigation into their potential chiral character. Skyrmions have previously been reported in FGT,\cite{Ding2020, Gatel2022, Yang2020} and the interfacial symmetry breaking introduced by CrSBr may play a pivotal role in stabilizing or modifying such topological textures, offering a novel pathway for engineering interfacial skyrmionics. Furthermore, recent proposals suggest that lateral exchange bias—induced via step-edge engineering—could provide a tunable degree of control over the N\'eel vector in CrSBr,\cite{pellet2025lateral} enabling new strategies for domain manipulation in two-dimensional magnetic heterostructures. These insights underscore the potential of CrSBr-based systems as a versatile platform for advancing the understanding and control of topological spin structures in vdW materials.

\section{Conclusion}

In summary, we demonstrate asymmetric magnetization switching and robust and tunable exchange bias in CrSBr/Fe\(_3\)GeTe\(_2\) vdW heterostructures, observed up to \(\sim130\,\)K. This orthogonally coupled system exhibits highly asymmetric hysteresis loops, characterized by abrupt monodomain switching on one side and gradual, domain-mediated reversal on the other. Depending on the applied field strength and polarity, the strength of the exchange bias can be switched. The origin of the change of the exchange bias is revealed by cross-sectional magnetic imaging, showing that the interplay of orthogonal anisotropies and interfacial exchange coupling leads to the formation of a complex 3D domain structure in FGT and canted magnetization in CrSBr. These complex domain structures give rise to the observed asymmetry in the magnetic hysteresis and provide a phenomenological framework for understanding exchange bias in orthogonally coupled systems.

Together, these findings point to a mechanism for stabilizing three-dimensional spin textures via symmetry breaking in layered AFM/FM heterostructures. The ability to manipulate such confined magnetic states through interfacial coupling opens promising directions for topology-driven spintronic devices, where precise control over nanoscale domain architectures is essential. These results also open new opportunities for the design of spintronic systems with tunable and asymmetric magnetic switching responses, which could be potentially useful for implementing energy-efficient spintronic devices.


\section{Experimental Section}
\subsection{Materials and Device Fabrication}

Bulk single crystals of CrSBr were purchased from HQ Graphene.\cite{hqgrapheneCrSBr} Bulk crystals of Fe\(_3\)GeTe\(_2\) (FGT) were grown using chemical vapor transport (CVT).\cite{deiseroth2006fe3gete2} Both the individual crystals and the assembled CrSBr/FGT vdW heterostructures were characterized using Raman spectroscopy. The measured spectra are in good agreement with reported Raman modes for CrSBr\cite{mondal2025raman} and Fe\(_3\)GeTe\(_2\)\cite{kong2021thickness} (see Figure~SI1b in the \textit{Supporting Information}).

Exfoliation of both crystals was carried out inside a glovebox with O\(_2\) and H\(_2\)O levels maintained below 0.5\,ppm. Few-layer flakes of Fe\(_3\)GeTe\(_2\) and CrSBr were exfoliated onto a Si/SiO\(_2\) (300\,nm) substrate and selected using optical microscopy. A dry transfer technique assisted by polydimethylsiloxane (PDMS) and polymethylmethacrylate (PMMA) was used to assemble the heterostructures in the desired sequence, ensuring clean and residue-free interfaces.

To protect the heterostructure from oxidation during brief air exposure while wire-bonding, an h-BN flake was transferred on top as a capping layer. Atomic force microscopy (AFM) was used to measure the thicknesses of the individual CrSBr and FGT layers in the heterostructures (refer to Figure~SI2 in the \textit{Supporting Information}).

\subsection{Magneto-transport measurements}

The h-BN-capped CrSBr/Fe\(_3\)GeTe\(_2\) heterostructures are wire-bonded and immediately loaded into a variable temperature insert (VTI) cryostat for magnetotransport measurements. The cryostat allows for the application of magnetic fields up to 12\,T. Since CrSBr is a semiconducting antiferromagnet, it is assumed that the electrical current flows exclusively through the metallic Fe\(_3\)GeTe\(_2\) layer.

A current of 50\,$\mu$A (corresponding to a current density of \(6 \times 10^9\,\textrm{A/m}^2\)) is applied along the \(x\)-direction of the Fe\(_3\)GeTe\(_2\) flake, and the anomalous Hall voltage (\(V_{xy}\)) is measured across transverse terminals aligned along the \(y\)-direction. During both the field-cooling and field-sweeping procedures, the magnetic field is applied in the OOP direction. A Keithley 2400 source meter supplies the current, and a Keithley 2182A nanovoltmeter is used to measure the transverse voltage.

\subsection{DC SQUID Magnetometry Measurements}
Magnetic hysteresis loops (M vs. H) and temperature-dependent magnetization curves (M vs. T) were acquired for a bulk CrSBr crystal using a Quantum Design SQUID MPMS3 Magnetometer. The M vs. H measurements involved sweeping the magnetic field in both IP and OOP orientations relative to the crystal's bulk structure.

\subsection{Scanning transmission electron microscopy (STEM) and energy dispersive X-ray (EDX) measurements:}
An electron-transparent cross-section lamella of the sample was prepared using a Ga\textsuperscript{+} focused ion beam and scanning electron microscope (FIB-SEM) FEI Helios dual-beam platform.
Scanning transmission electron microscopy (STEM) was carried out using a TFS Spectra 300 TEM equipped with a Schottky field emission gun operated at 300\,kV, a CEOS probe aberration corrector, a high-angle annular dark-field (HAADF) detector, and an EDX Super-X detector.

\subsection{Off-axis electron holography measurements:}
 
Off-axis electron holography was carried out using a TFS Titan TEM equipped with a Schottky field emission gun operated at 300\,kV, a CEOS image aberration corrector, a field-free Lorentz mode, an electron biprism, and a 2k x 2k Gatan CCD camera.\cite{Thust2016} A Gatan liquid-nitrogen-cooled specimen holder (model 636) and a temperature controller (model 613-0500) were used to perform experiments at low temperature. In Lorentz mode, the objective lens was used to apply external magnetic fields to the sample, which were precalibrated using a Hall sensor. 
In off-axis electron holography, the interference of a wave passing through the sample (object wave) and a wave traveling in vacuum (reference wave) is recorded on a camera. The hologram is then processed using Fourier transform operations to recover the phase, which is sensitive to electromagnetic fields. Here, phase images were reconstructed using the Digital Micrograph software (Gatan) with the Holoworks plugin.\cite{Voelkl1995} Color-coded maps that show the direction of the B field were obtained from the gradient of the phase images.

\medskip
\textbf{\textit{Supporting Information}} \par 
Supporting Information is available from the Wiley Online Library or from the authors.

\medskip
\textbf{Acknowledgements} \par 
We gratefully acknowledge Prof. Dr. Bettina V. Lotsch for providing the bulk Fe\(_3\)GeTe\(_2\) crystals. We acknowledge funding from Alexander von Humboldt Foundation for the Humboldt Postdoctoral Fellowship (Grant number: Ref 3.5-IND-1216986-HFST-P), EU Marie-Curie Postdoctoral Fellowship ExBiaVdW (Grant ID: 101068014), Deutsche Forschungsgemeinschaft (DFG, German Research Foundation) – Spin$+$X TRR 173–268565370 (Projects No. A01, A03, A11, B02, B15 and A12), Deutsche Forschungsgemeinschaft (DFG, German Research Foundation) – ELASTO-Q-MAT TRR 288 – 422213477 (Projects No. A09 and A12), DFG Project No. 358671374, Graduate School of Excellence Materials Science in Mainz (MAINZ) GSC 266, the MaHoJeRo (DAAD Spintronics network, Projects No. 57334897 and No. 57524834), the Research Council of Norway (Centre for Quantum Spintronics - QuSpin No. 262633), The European Union’s Horizon 2020 Research and Innovation Programme under grant agreement 856538 (project “3D MAGIC”). 

\medskip

%
\bibliographystyle{MSP}
\bibliography{references}

\newpage

\begin{figure}
\centering
\includegraphics[width=\linewidth]{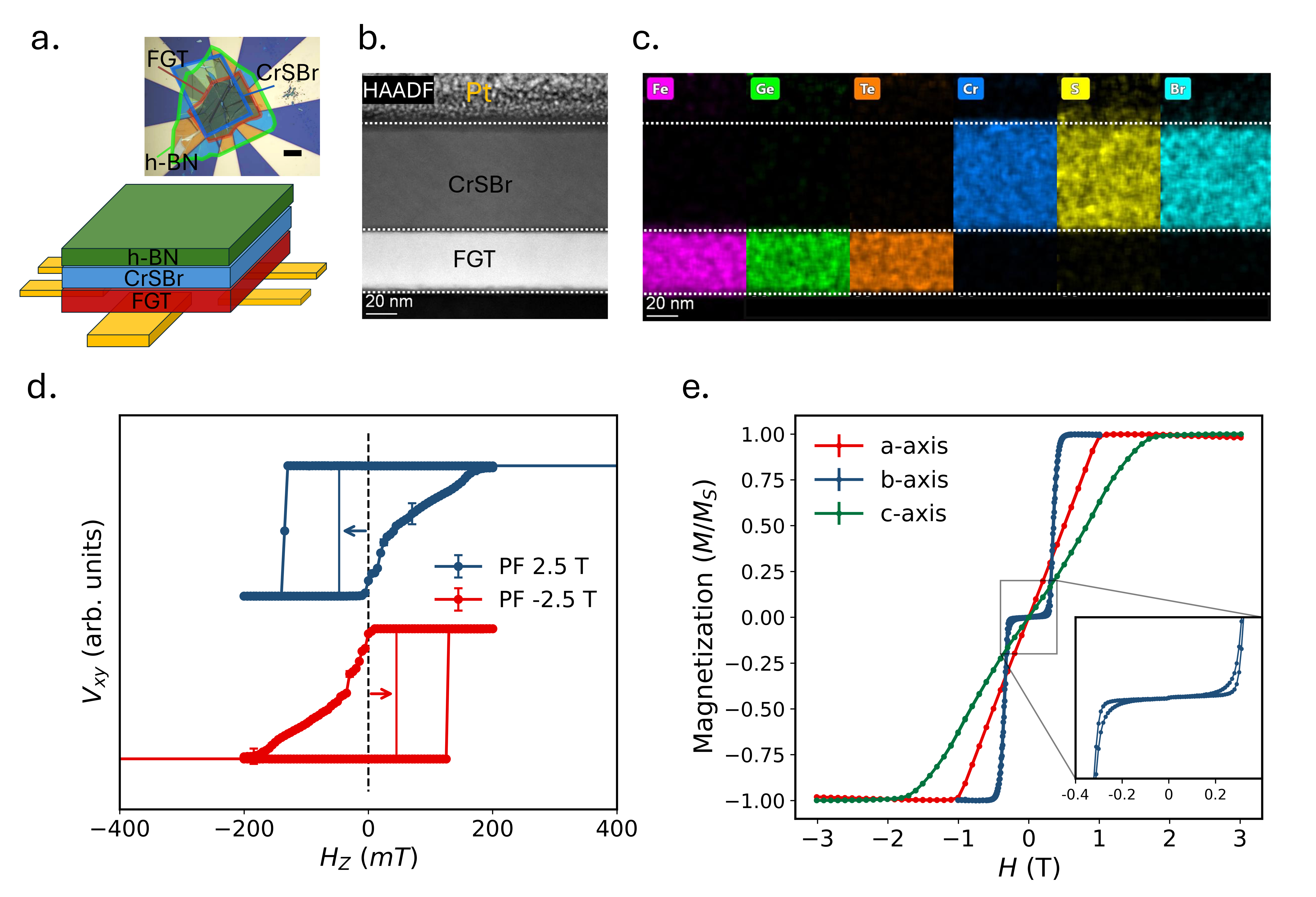}
\caption{\textbf{Asymmetric switching in a CrSBr/FGT vdW heterostructure.}
(a) A schematic diagram of freestanding flakes on pre-patterned AHE contacts is shown, with an optical micrograph in the inset (scale bar: 10\,$\mu$m).
(b) The STEM image shows the cross-section of the heterostructure. 
(c) The corresponding EDX elemental maps indicate that there is no interdiffusion of the constituent elements, suggesting a pristine interface.
(d) Asymmetric switching occurs at 10\,K for preset-fields of $+2.5$\,T (blue) and $-2.5$\,T (red). 
(e) Magnetization of CrSBr as a function of applied field along the symmetry axes is displayed. The inset shows spin flip of CrSBr along its easy axis.
}

\label{fig:device_EB_SQUID}
\end{figure}

\begin{figure}
\centering
\includegraphics[width=\linewidth]{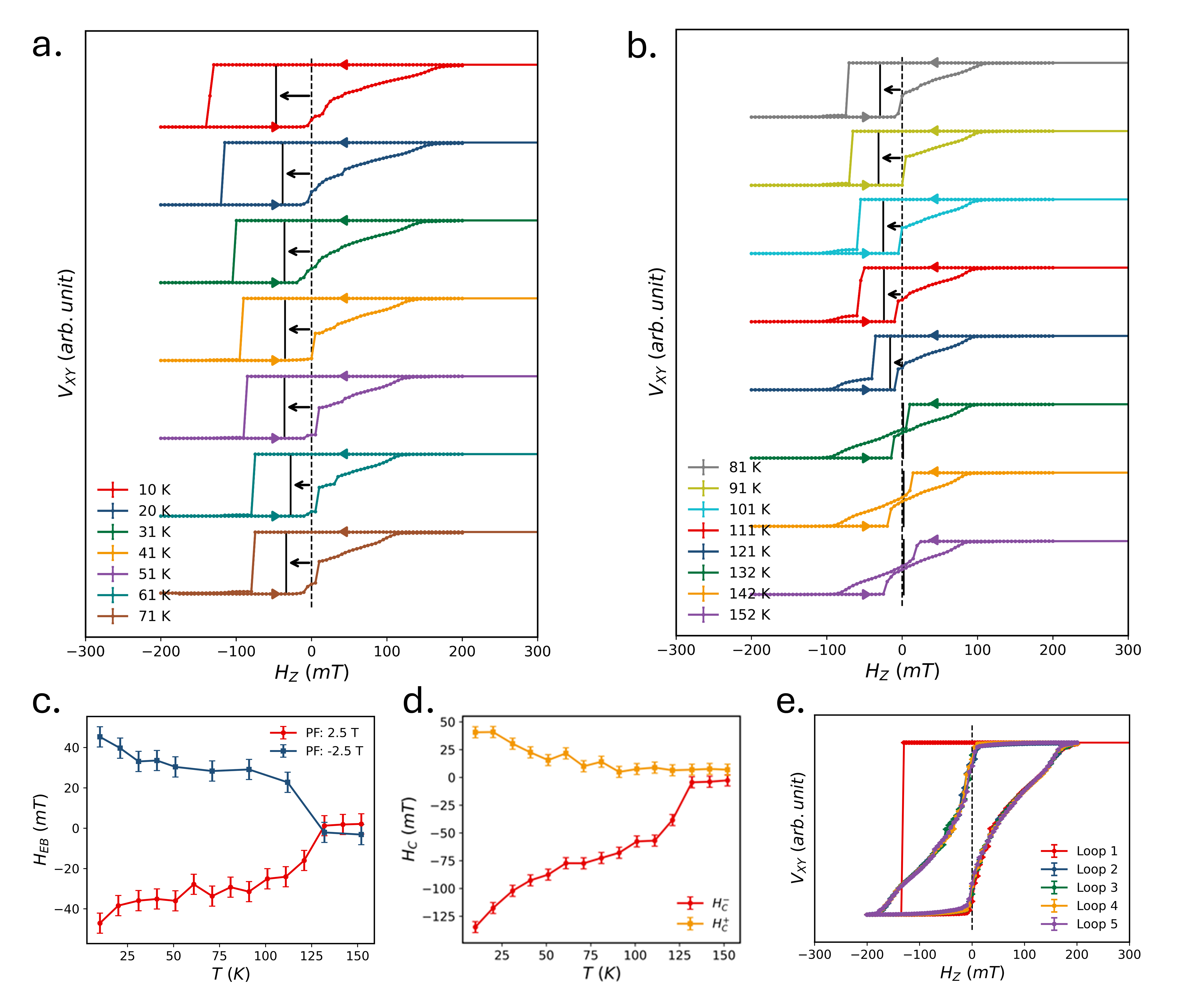}
\caption{\textbf{Characteristics of asymmetric switching: temperature dependence and training effect.} 
(a,b) Hall effect transport data for a preset-field of $+2.5$\,T at various temperatures from 10\,K to 152\,K are shown. The characteristic asymmetric shape of the hysteresis persists up to 132\,K where the exchange bias diminishes to zero. The error bars are shown but are smaller than the linewidth. 
(c) Exchange bias is presented as a function of temperature for preset-fields of $+2.5$\,T (red) and $-2.5$\,T (blue). The blocking temperature of exchange bias aligns with the 132\,K $T_N$ of CrSBr.
(d) Positive (yellow) and negative (red) switching fields are plotted as a function of temperature for a preset-field of $+2.5$\,T, describing the asymmetric shape of the hysteresis loop. 
(e) Asymmetry is evident only in the first field sweep (red) following the preset condition and disappears in subsequent cycles.}

\label{fig:fig2}
\end{figure}

\begin{figure}
\centering
\includegraphics[width=\linewidth]{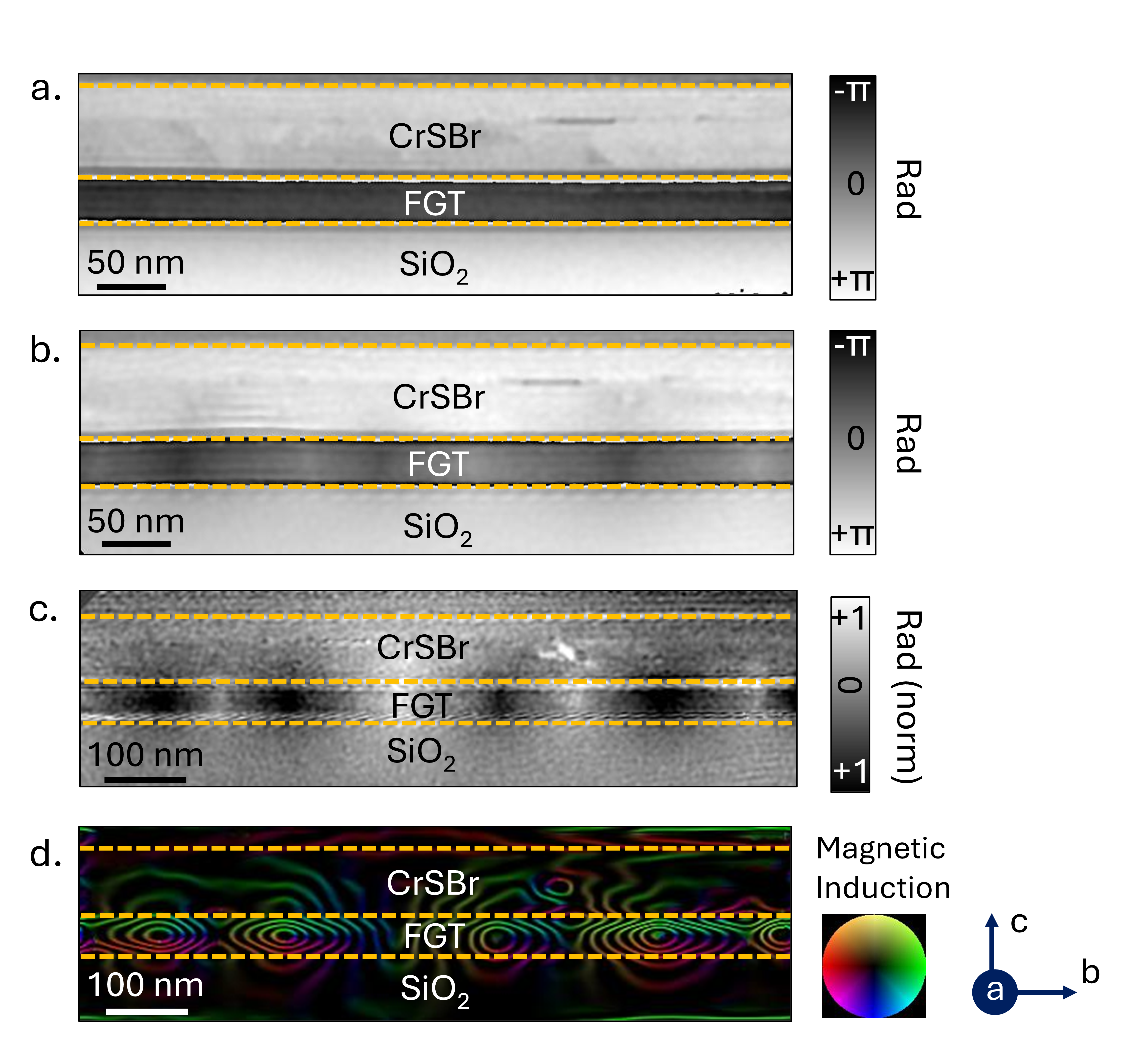}
\caption{\textbf{Electron Holography images of the CrSBr/FGT heterostructure along the $bc$ plane measured at 0 mT field and 95 K.} (a) Phase map measured at zero field after application of a 1\,T field along the $c$ axis. (b) Phase map of the demagnetized state.  (c) Magnetic phase image of the zero-field cooled state after subtraction of the nonmagnetic contribution using a phase image acquired above the $T_C$ of FGT. (d) Color-coded phase gradient map that shows the direction of the magnetic induction field B with a contour spacing of 2$\pi$/30.
}
\label{fig:fig3}
\end{figure}

\begin{figure}
\centering
\includegraphics[width=0.85\linewidth]{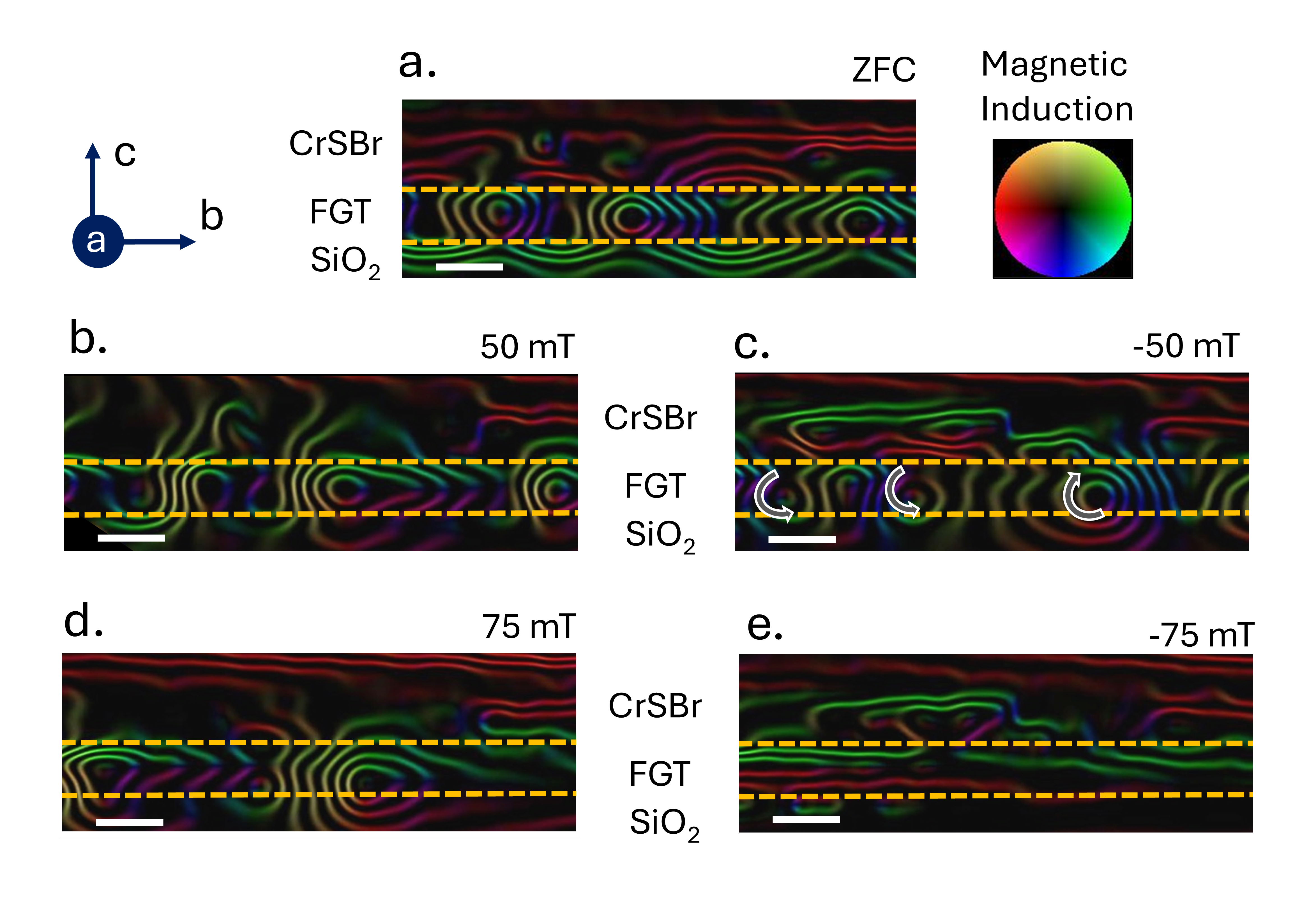}
\caption{\textbf{History-dependent field imaging.}
Magnetic induction maps acquired at 95\,K and zero applied field after sequential OOP magnetic field application. Images a--e were recorded chronologically: a. zero-field-cooled state; b. after +50\,mT OOP field; c. after $-50$\,mT OOP field; d. after +75\,mT OOP field; e. after $-75$\,mT OOP field. Scale bar: 50\,nm.
}

\label{fig:fig4}
\end{figure}

\begin{figure}
\centering
\includegraphics[width=\linewidth]{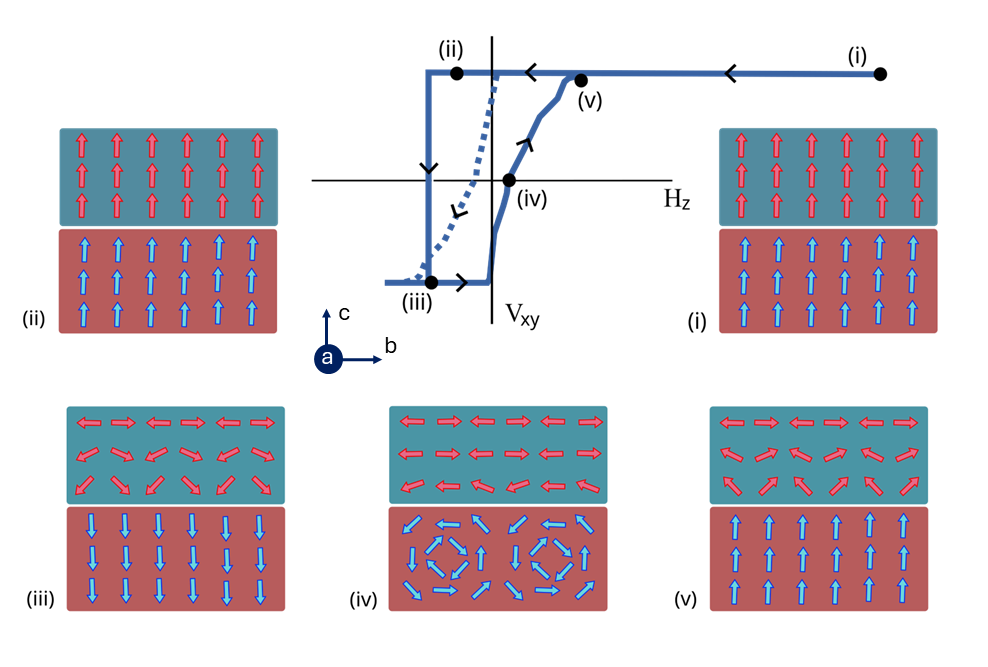}
\caption{\textbf{Schematic model for asymmetric switching.} 
(i) Application of a 2.5\,T OOP preset-field saturates CrSBr and FGT in the OOP direction. (ii) Once the preset-field is removed, the adjacent layers of FGT and CrSBr on either side of the interface remain aligned due to exchange coupling. (iii) When a sufficiently strong negative magnetic field is applied, the bulk FGT layer undergoes magnetization reversal. However, the proximitized layers of FGT and CrSBr go to a canted state because of the IP anisotropy of CrSBr. The antiferromagnetically coupled spins depicted in the same row in panels (iii) and (v) illustrate the interlayer exchange interaction. They are drawn in the same row for better legibility of the schematic. (iv) The magnetization reversal from negative to positive side is gradual due to domain nucleation. (v) As FGT saturates again in the positive direction, the magnetization in CrSBr forms a canted state rather than fully aligning along the OOP direction. The dashed hysteresis loop shown in hysteresis corresponds to subsequent measurement cycles following the initial loop. These hysteresis loops cycle between steps (iii), (iv), and (v).
}

\label{fig:fig5}
\end{figure}



\end{justify}
\end{document}